\documentclass{rmf-d}
\usepackage{nopageno,rmfbib,multicol,times,epsf,amsmath,amssymb,cite}
\usepackage[latin1]{inputenc}
\usepackage[]{caption2}
\usepackage{graphics}
\usepackage{graphicx}
\def\rmfcornisa{Research OR Education of Physics \hfill\rmf\ {\bf ??} (*?*) ???--???
\hfill MES? A\~NO?}

\def\rmfcintilla{{\it Rev.\ Mex.\ Fis.\/} {\bf ??} (*?*) (????) ???--???}
\clearpage \rmfcaptionstyle 
\begin{document}
\title{Cluster Reducibility of Multiquark Operators\\and
Tetraquark-Adequate QCD Sum Rules\vspace{-6pt}}
\author{Wolfgang Lucha}\address{Institute for High Energy Physics,
Austrian Academy of Sciences, Nikolsdorfergasse 18, A-1050 Vienna,
Austria\and e-mail: Wolfgang.Lucha@oeaw.ac.at }
\author{Dmitri Melikhov}\address{D.~V.~Skobeltsyn Institute of
Nuclear Physics, M.~V.~Lomonosov Moscow State University, 119991
Moscow, Russia\and Joint Institute for Nuclear Research, 141980
Dubna, Russia\and Faculty of Physics, University of Vienna,
Boltzmanngasse 5, A-1090 Vienna, Austria\and e-mail:
dmitri\_\,melikhov@gmx.de}
\author{Hagop Sazdjian}\address{Universit\'e Paris-Saclay,
CNRS/IN2P3, IJCLab, 91405 Orsay, France\and e-mail:
sazdjian@ijclab.in2p3.fr}

\maketitle \recibido{day month year}{day month year
\vspace{-12pt}}

\begin{abstract}If connecting properties of a multiquark hadron
with those exhibited by its constituents, QCD sum rules inferred
along the routes of traditional wisdom necessarily involve
contributions not related at all to and thus not presenting
information about multiquarks. Realizing this deficiency, we
propose to increase, for the example of tetraquarks, the
predictive power of the QCD sum-rule formalism by disposal of all
of the unwanted contributions from the very beginning; this move
is easily accomplished by subjecting the contributions to our
perspicuous selection criterion.\end{abstract}

\keys{quark bound states, exotic hadron, multiquark, tetraquark,
QCD sum rules, interpolating operator, multiquark adequacy,
clustering\vspace{-4pt}}

\pacs{\bf{\textit{14.40.Rt, 12.38.Lg}}\vspace{-4pt}}
\begin{multicols}{2}

\section{Bound states induced by strong interactions}The most basic
description of all strong interactions observed in elementary
particle physics is (at least, at present) provided by
\emph{quantum chromodynamics} (QCD), a quantum field theory,
exhibiting invariance under local transformations forming the
non-Abelian gauge group SU(3). Its basic degrees of~freedom
encompass the gluons (vector bosons transforming according to the
eight-dimensional adjoint representation of SU(3)) plus a set of
flavoured quarks (fermions transforming according to the
fundamental, three-dimensional representation of~SU(3)). For both
quarks and gluons, the degree of freedom induced by this local
SU(3) gauge invariance is referred to as their~colour.

Viewed from the angle of QCD, a hadron is perceived as a bound
state of the quark and gluon degrees of freedom. Thus, for any
hadron $H$ an interpolating operator ${\cal O}$ of $H$ is a
(local) \emph{gauge-invariant} operator composed of quark and
gluon fields that has a nonzero overlap with the hadron's state
$|H\rangle$, that is, enjoys a nonvanishing matrix element if
sandwiched between the QCD vacuum $|0\rangle$ and the hadron state
$|H\rangle$: $\langle0|{\cal O}|H\rangle\ne0$.

Recently, we worked out
\cite{LMS1,LMS2,LMS3,LMS4,LMS5,LMS6,LMS7,LMS8,LMS9,LMS10,LMS11}
for a popular approach to hadrons called QCD sum rules a, from our
point of view, kind of improvement tailored to the peculiarities
of \emph{exotic}~hadrons.

\emph{QCD sum rules} \cite{SVZ} relate observable features of
hadrons to the degrees of freedom (quarks and gluons) and
parameters of the quantum field theory governing the strong
interactions, quantum chromodynamics; they may be easily
established by evaluation of suitable $n$-point correlation
functions of hadron interpolating operators (defined by quarks and
gluons) at both hadron level (by inserting a complete set of
hadron states) and QCD level (by adopting the operator product
expansion \cite{KGW}). By construction, their distinctive
characteristic is to provide a \emph{nonperturbative} approach by
means of analytic relationships.

All QCD-controlled colour-singlet bound states of quarks and
gluons may be categorized into two disjoint sets, ordinary hadrons
comprising quark--antiquark mesons and three-quark baryons, and
\emph{exotic hadrons}, with all \emph{multiquarks} (containing as
constituents more quarks than merely one quark--antiquark pair or
just three quarks) as a prominent set of representatives.

The most distinctive feature common to all multiquarks is their
ability for clustering \cite{LMS7}: in contrast to ordinary
hadrons, every multiquark may decompose into clusters of,
eventually, \emph{ordinary} hadrons. As a consequence, any
multiquark must be viewed simultaneously as a strongly bound
compact state and as a rather loosely bound aggregate of these
ordinary hadrons.

For the sake of clarity, our ideas are best illustrated for the
least complex systems presumably realized by multiquarks of
smallest possible number of constituents: bound states of two
quarks and two antiquarks, forming the subset of
\emph{tetraquarks}.

\section{Multiquark-hadron interpolating operators}The starting
point of such line of argument is the specification of any adopted
multiquark interpolating operator with respect to its for our
analysis most useful internal Lorentz, colour and flavour
structure in terms of quark fields $q^\alpha_a(x)$ endowed with
colour indices ($\alpha,\beta,\dots$) as well as flavour indices
($a,b,\dots$).

Concerning notations, we find it preferable to skip all (for all
following considerations irrelevant) references to parity or spin
degrees of freedom, and thus suppress all Dirac matrices.

\subsection{Tetraquark interpolating operators: colour singlets}
All hadrons are colour singlets, vulgo colourless. Thus, every
admissible hadron interpolating operator ${\cal O}$ has to be
likewise an overall colour singlet. This constraint, however, does
in no way predetermine the internal colour structure of the
operator ${\cal O}$. For tetraquark interpolating operators
$\theta$, this trivial insight entails that their colour
composition can be chosen to be, e.g., of the singlet--singlet
nature $\theta\sim(\overline q_\alpha\,q^\alpha)\,(\overline
q_\beta\,q^\beta)$, or, in terms of the Levi-Civita symbol
$\epsilon_{\alpha\beta\gamma}$, of the antitriplet--triplet form
$\theta\sim\epsilon_{\alpha\beta\gamma}\,(q^\beta\,q^\gamma)\,
\epsilon^{\alpha\delta\varepsilon}\,(\overline q_\delta\,\overline
q_\varepsilon)$, or, adopting all generators $T^A$,
$A=1,2,\dots,8$, of the gauge group SU(3) in the fundamental
representation, of the octet--octet type $\theta\sim(\overline
q\,T^A\,q)\,(\overline q\,T^A\,q)$. By means of a Fierz
transformation \cite{MF}, however, any colour structure of
tetraquark interpolating operators can be cast into a sum of
products of two colour-singlet quark bilinears; these findings
clearly relativize the significance of an ``appropriate'' choice
of interpolating operator. That is to say, all choices are
equivalent, it suffices to consider the singlet--singlet
structure.

\subsection{Tetraquark interpolating operators: quark flavours}In
principle, as a consequence of the clustering property of all
multiquarks, the flavour composition of these is subject to the
same constraints as those of ordinary hadrons, that is,~to none.
Table 1 of Ref.~\cite{LMS4} provides an exhaustive classification
of the potential open or hidden quark-flavour content
of~tetraquarks.

For tetraquarks, upon taking advantage of the opportunity to
perform Fierz transformations, it clearly suffices to confine all
investigations to generic tetraquark interpolating
operators\begin{equation}\theta_{\bar ab\bar cd}(x)\equiv j_{\bar
ab}(x)\,j_{\bar cd}(x)\label{tio}\end{equation}defined by a
product of two colourless quark-bilinear
currents\begin{equation}j_{\bar ab}(x)\equiv\overline
q_{a\alpha}(x)\,q_b^\alpha(x)\ .\label{qbc}\end{equation}

An evidently hindering implication \cite{SC,SW} of multiquark
clustering is that two-point correlation functions of tetraquark
interpolating operators (\ref{tio}) receive not merely
nonfactorizable (NF) contributions, that potentially convey
information about tetraquarks, but also factorizable terms, that
definitely do not:\begin{align*}&\langle{\cal T}(\theta_{\bar
ab\bar cd}(x)\, \theta^\dag_{\bar ab\bar
cd}(0))\rangle=\langle{\cal T}(\theta_{\bar ab\bar
cd}(x)\,\theta^\dag_{\bar ab\bar cd}(0))\rangle_{\rm
NF}\nonumber\\&+\langle{\cal T}(j_{\bar ab}(x)\,j^\dag_{\bar
ab}(0))\rangle\,\langle{\cal T}(j_{\bar cd}(x)\,j^\dag_{\bar
cd}(0))\rangle\ .\end{align*}

With respect to the tetraquark flavour degrees of freedom, in the
following we shall focus, for definiteness, to the case of
\emph{flavour-exotic} tetraquark hadrons: bound states of two
quarks and two antiquarks that carry four mutually different
flavours, $a\ne b\ne c\ne d$. For these, precisely two linearly
independent interpolating operators do exist, in representation
(\ref{tio})~given~by$$\theta^{(1)}_{\bar ab\bar cd}(x)\equiv
j_{\bar ab}(x)\,j_{\bar cd}(x)\ ,\qquad\theta^{(2)}_{\bar ad\bar
cb}(x)\equiv j_{\bar ad}(x)\,j_{\bar cb}(x)\ .$$

\section{Correlation function of four quark bilinears}A very
promising approach fully in line with Eq.~(\ref{tio}) is to reap
all the information sought from tetraquark intermediate states of
correlation functions of four suitably chosen colour-singlet
quark-bilinear currents (interpolating only \emph{ordinary}
mesons).

In our search for \emph{multiquark adequacy} of any tool utilized
for the investigation of exotic hadrons, an (evidently) decisive
move is to find a means both to discard all those contributions
that definitely won't have an impact on the multiquarks under
\end{multicols}\begin{figure*}[hb]\centering
\includegraphics[width=.3764\textwidth]{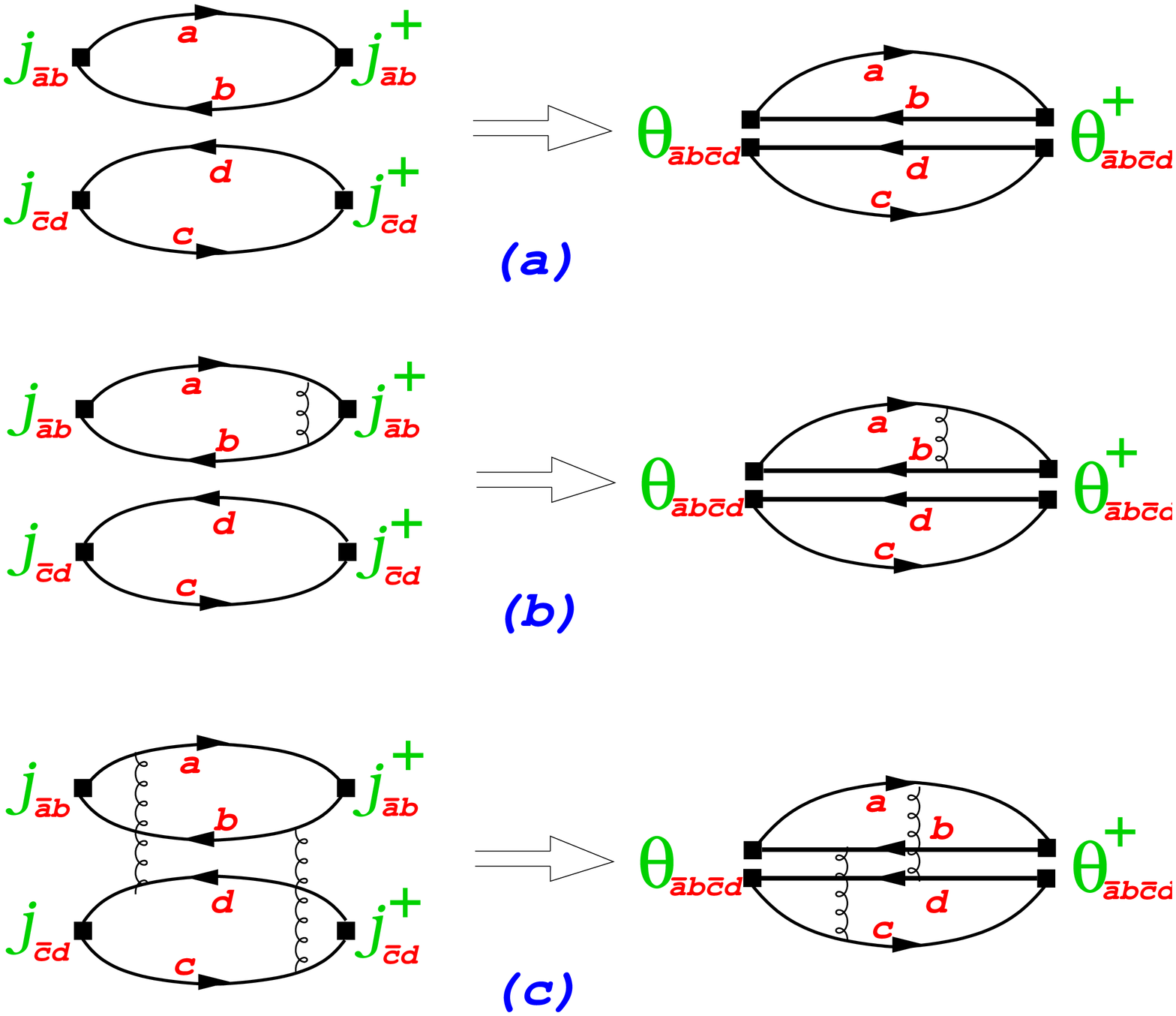}\hspace{10ex}
\includegraphics[width=.3764\textwidth]{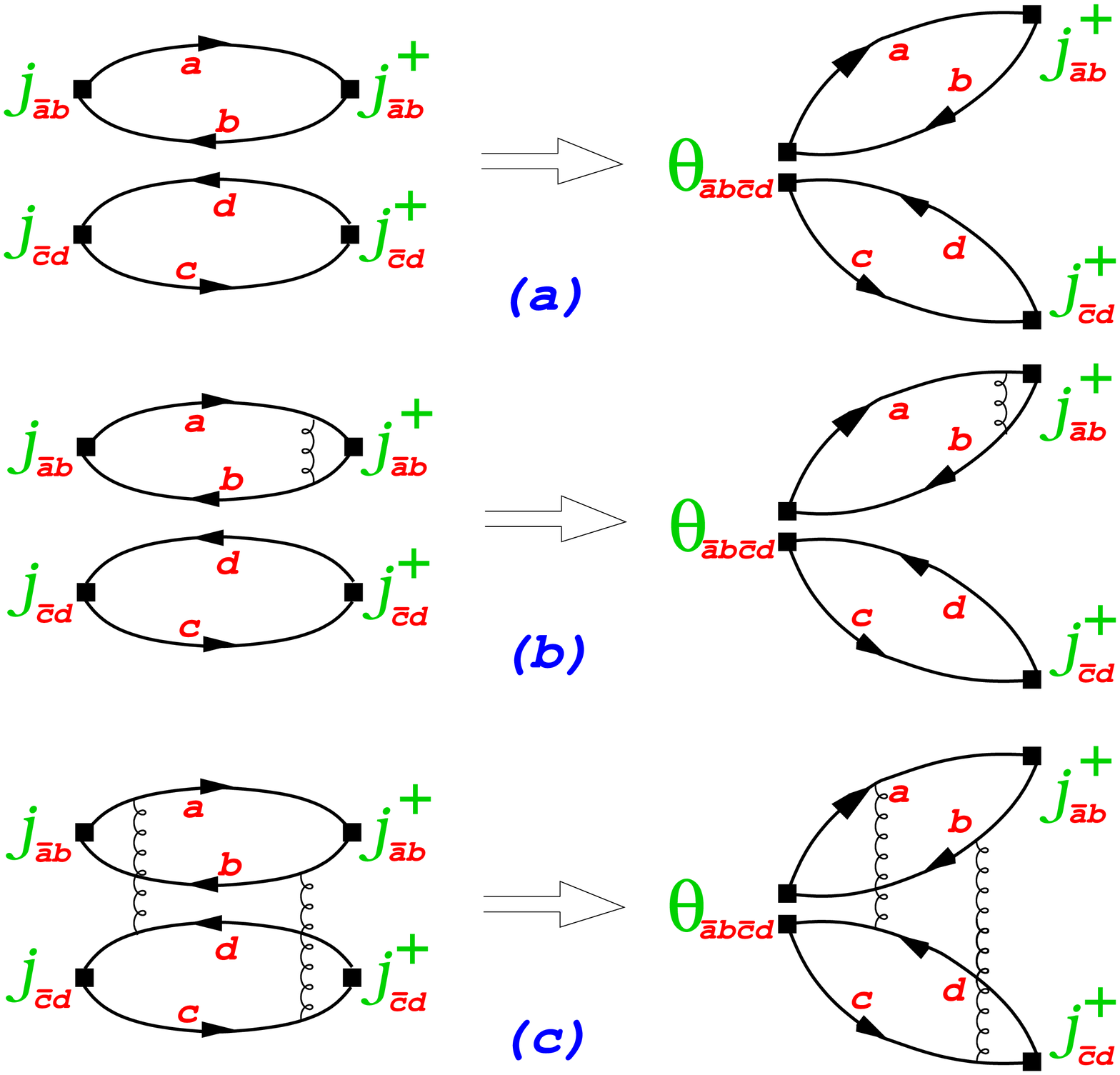}
\caption{Exemplary perturbative contributions of lowest
conceivable strong-coupling orders $\alpha_{\rm s}^0$ (a),
$\alpha_{\rm s}$ (b), and $\alpha_{\rm s}^2$ (c) to
\emph{flavour-preserving correlation functions} of four
quark-bilinear currents $j$ and the latters' configuration-space
vertex contractions to correlation functions of either \emph{two}
tetraquark interpolating operators $\theta^{(1)}$ (left)
\cite{LMS6,LMS9}, or \emph{one} tetraquark interpolating operator
$\theta^{(1)}$ plus two quark-bilinear currents~(right)
\cite{LMS6}.}\label{Fig.34}\end{figure*}\begin{multicols}{2}
\end{multicols}\begin{figure*}[ht]\centering
\includegraphics[width=.3764\textwidth]{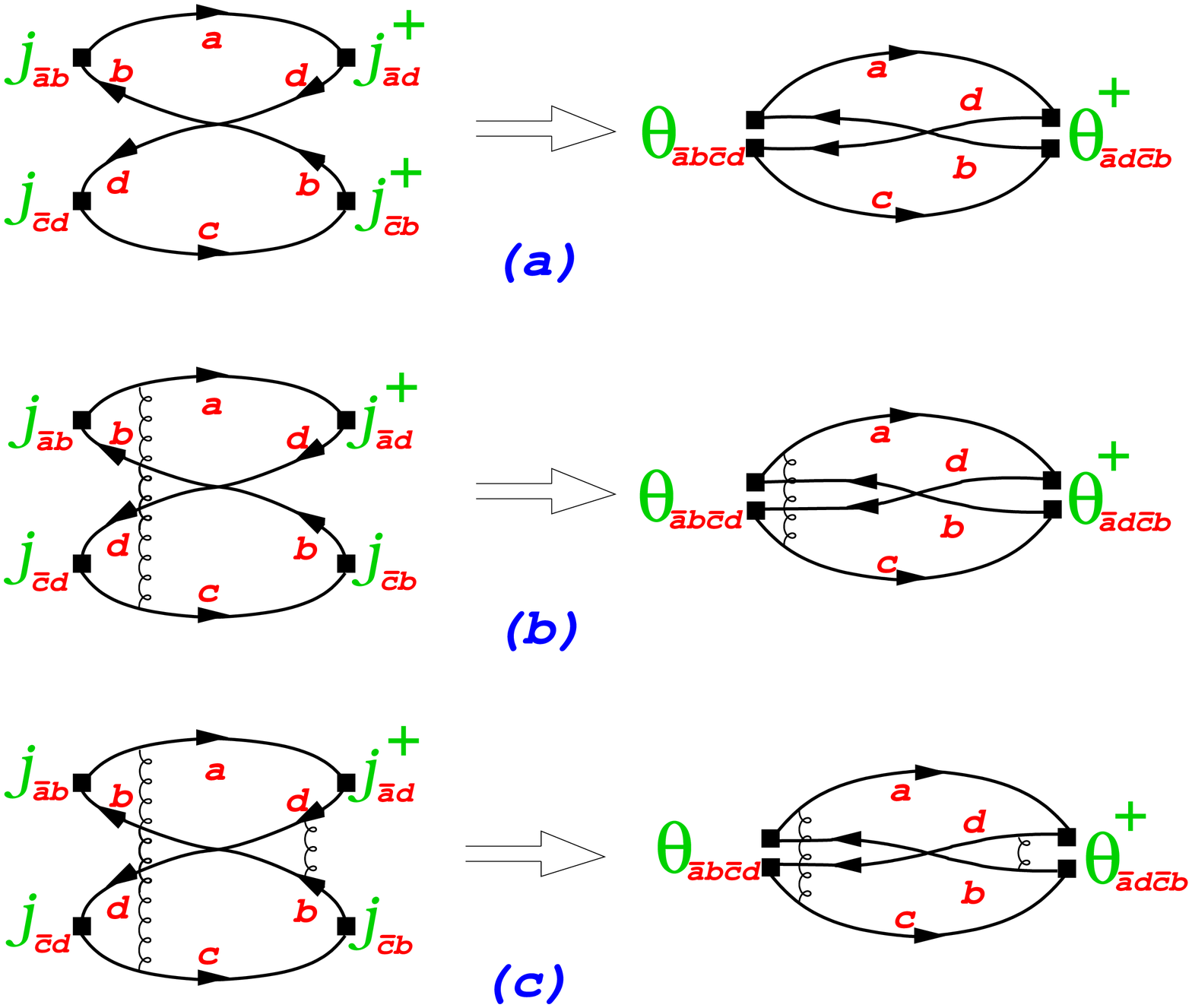}\hspace{10ex}
\includegraphics[width=.3764\textwidth]{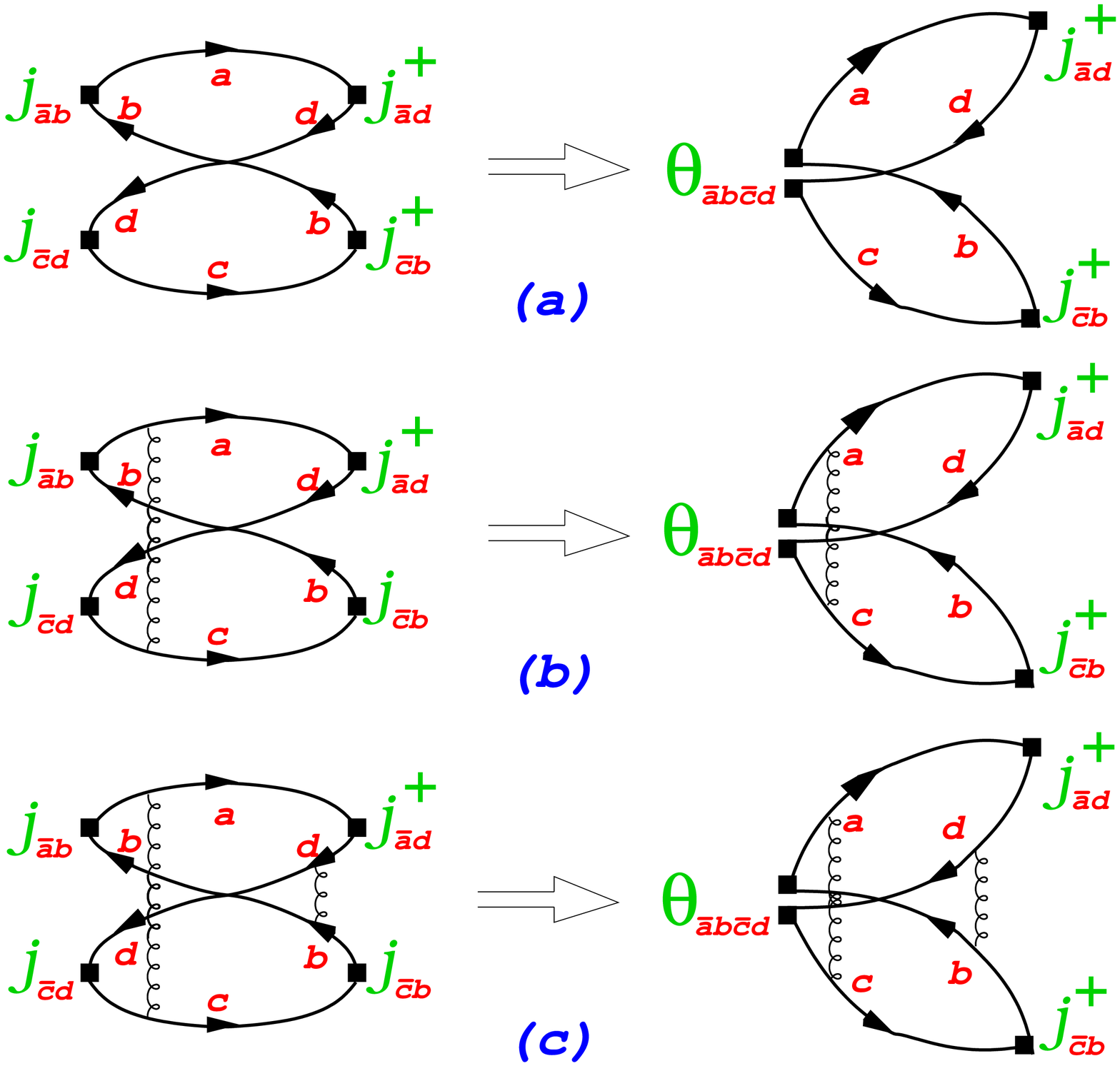}
\caption{Exemplary perturbative contributions of lowest
conceivable strong-coupling orders $\alpha_{\rm s}^0$ (a),
$\alpha_{\rm s}$ (b), and $\alpha_{\rm s}^2$ (c) to
\emph{flavour-reordering correlation functions} of four
quark-bilinear currents, and configuration-space vertex
contractions thereof, to correlation functions of either
\emph{two} tetraquark interpolating operators $\theta^{(1)}$,
$\theta^{(2)}$ (left) \cite{LMS6,LMS8} or \emph{one} tetraquark
interpolating operator $\theta^{(1)}$ and two quark-bilinear
currents~(right) \cite{LMS8}.}\label{Fig.1213}\end{figure*}
\begin{multicols}{2}
\noindent study and to identify and retain exactly all those
contributions --- labelled \emph{multiquark-phile} --- that
\emph{might} be of relevance for the description of any
multiquarks in the focus of our interest. For the particular case
of tetraquarks, we formulated, in terms of Feynman diagrams, a
criterion \cite{LMS1} enabling us to distil only the
\emph{tetraquark-phile} \cite{LMS2,LMS5} among the entirety of
contributions to a four-point correlation function of
quark-bilinear currents:\begin{quote}In terms of the Mandelstam
variable $s$ defined by the four relevant external momenta $p_1$,
$p_2$, $p_3$, $p_4$ on equal footing via
$s\equiv(p_1+p_2)^2=(p_3+p_4)^2$, a \emph{tetraquark-phile}
\cite{LMS2,LMS5} Feynman diagram has to $\bullet$ exhibit a
non-polynomial dependence on $s$ and $\bullet$ develop a branch
cut, defined by a branch point $\hat s$ governed by the masses
$m_a$, $m_b$, $m_c$, $m_d$ of the (anti)quarks $\overline q_a$,
$q_b$, $\overline q_c$, $q_d$ forming the \emph{tetraquark hadron}
according to $\hat s=(m_a+m_b+m_c+m_d)^2.$\end{quote}The (really
pivotal) question whether some Feynman diagram under consideration
is capable of developing such four-quark singularities may be
unambiguously decided, by means of the Landau equations \cite{LEq}
(for explicitly worked out examples of the latters' application to
\emph{tetraquarks}, consult Refs.~\cite{LMS3,LMS8,LMS11}). The
fulfilment of this necessary but not sufficient prerequisite
guarantees straightforwardly a Feynman diagram's
\emph{suitability} to contribute to the formation of the suspected
tetraquark pole (located, for a generic tetraquark $T$ of mass
$M_T$, at~$s=M_T^2$).

Flavour exoticism characterizing the subset of tetraquarks
presently in our focus prompts us to categorize the considered
four-point correlation functions of quark-bilinear currents
(\ref{qbc}) into \emph{flavour-preserving} (occasionally dubbed
direct) ones and \emph{flavour-reordering} (occasionally called
recombination) ones:\begin{itemize}\item flavour-preserving
correlation functions are of the form\begin{equation}\langle{\cal
T}(j_{\bar ab}\,j_{\bar cd}\,j^\dag_{\bar ab}\,j^\dag_{\bar
cd})\rangle\ ,\qquad\langle{\cal T}(j_{\bar ad}\,j_{\bar
cb}\,j^\dag_{\bar ad}\,j^\dag_{\bar cb})\rangle\
;\label{fpC}\end{equation}\item flavour-reordering correlation
functions are of the form\begin{equation}\langle{\cal T}(j_{\bar
ad}\,j_{\bar cb}\,j^\dag_{\bar ab}\,j^\dag_{\bar cd})\rangle\
.\label{frC}\end{equation}\end{itemize}For such categories,
Figs.~\ref{Fig.34} and \ref{Fig.1213}, respectively, exemplify the
perturbative contributions of lowest orders in the coupling
$\alpha_{\rm s}$, related to the fundamental coupling parameter of
QCD, $g_{\rm s}$, by$$\alpha_{\rm s}\equiv\frac{g_{\rm
s}^2}{4\pi}\ .$$Nonperturbative contributions involving vacuum
condensates --- whose presence is required by these correlation
functions' operator product expansions and below implicitly
understood --- may be and have been investigated on equal footing
\cite{LMS8,LMS10}.

Finally, pairwise configuration-space coordinate merging of
quark-bilinear operators reduces their four-point functions,
\begin{itemize}\item upon a twofold contraction, to the correlation
functions of two tetraquark interpolating operators
(Figs.~\ref{Fig.34} and \ref{Fig.1213}, left), among others
yielding the tetraquark masses $M_T$;\item upon single
contractions, to the correlation functions of just a sole
tetraquark interpolating operator and the two unaffected
quark-bilinear currents (Figs.~\ref{Fig.34} and \ref{Fig.1213},
right) that govern, for instance, the tetraquarks'
decay~widths.\end{itemize}

Application of our criterion \cite{LMS1} shows (after unfolding
the Feynman diagrams in the flavour-reordering case (\ref{frC}))
without any difficulty that, \emph{exclusively}, contributions of,
at least, order $\alpha_{\rm s}^2$ (Figs.~\ref{Fig.34}(c) and
\ref{Fig.1213}(c)) may support four-quark singularities,
contributing to the eventual development of a tetraquark pole.

\section{Yield: multiquark-adequate QCD sum rule}Ultimately,
allowing insights, at QCD level acquired, into the separation of
contributions to four-point correlation functions of
quark-bilinear currents into those that are multiquark-phile and
those that undoubtedly are not to enter the QCD sum-rule machinery
implicates \emph{QCD--hadron interrelations} \cite{LMS6,LMS8} more
adequate to the needs of multiquarks than the traditional ones.

\begin{figure}[ht]\centering
\includegraphics[width=.884\textwidth]{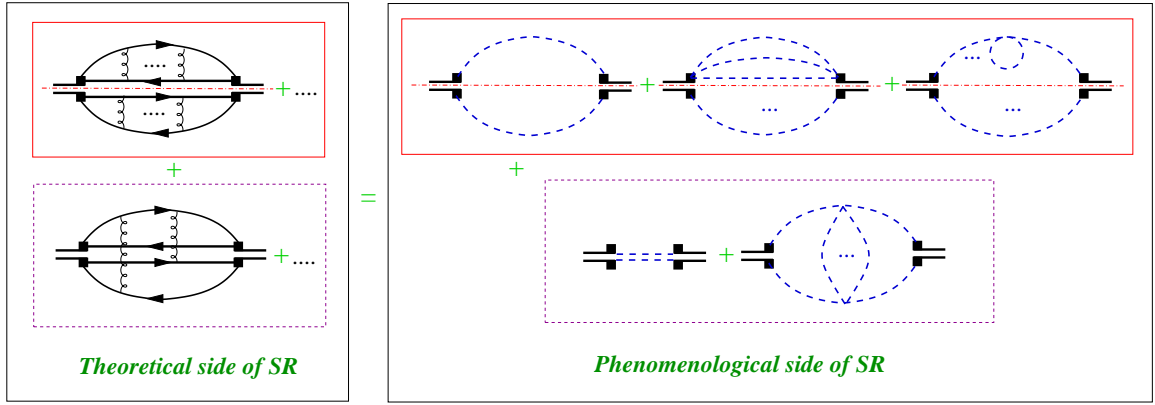}
\caption{Disentanglement of the QCD sum-rule output of
\emph{flavour-preserving} correlation functions of two tetraquark
interpolating operators $\theta$, generating pairs of
\emph{ordinary-meson} QCD sum rules reflecting the nature of the
\emph{factorizable} contributions (top row, separated by
dash-dotted lines) as well as the requested tetraquark-adequate
QCD sum rules embodying the totality of \emph{nonfactorizable}
contributions (bottom row) \cite{LMS6,LMS9}.}\label{Fig.5}
\end{figure}

\begin{figure}[htb]\centering
\includegraphics[width=.521\textwidth]{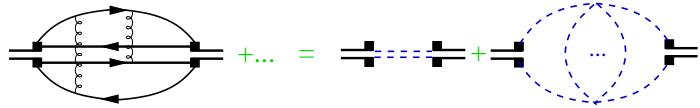}
\caption{Symbolic depiction of the tetraquark-adequate QCD sum
rules \cite{LMS6} as deduced from the \emph{flavour-preserving}
correlation functions of two tetraquark interpolating operators
$\theta$, relating exclusively \emph{tetraquark-phile} QCD-level
contributions [involving two or more internal gluons (curly
lines)] to hadron-level contributions of tetraquark poles (dashed
double lines) and nonfactorizable mesonic contributions (dashed
lines).}\label{Fig.7}\end{figure}

In the flavour-preserving case, this necessitates to identify and
discard a pair of ordinary-meson QCD sum rules (Fig.~\ref{Fig.5}),
to be left with \emph{tetraquark-focused} QCD sum rules
\cite{LMS6} (Fig.~\ref{Fig.7}).
\end{multicols}\begin{multicols}{2}

\begin{figure}[htb]\centering
\includegraphics[width=.48\textwidth]{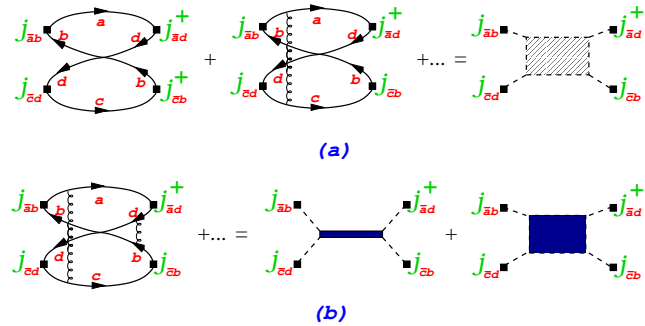}
\caption{Disjointedness of the two interrelationships inferable
from \emph{flavour-reordering} correlation functions of four
quark-bilinear currents \cite{LMS8}: (a) duality of QCD-level
contributions found to be \emph{not} tetraquark-phile and
hadron-level contributions not involving two-meson $s$-channel
cuts (hatched rectangle), on the one hand, and (b)
tetraquark-adequate QCD sum rules relating all
\emph{tetraquark-phile} QCD-level contributions to hadron-level
contributions exhibiting either tetraquark poles (horizontal bar)
or two-meson $s$-channel cuts (filled rectangle), on the
other~hand.}\label{Fig.14}\end{figure}

In the flavour-reordering case, things are not that easy
\cite{LMS8}. Jettisoning all clearly multiquark-irrelevant ballast
(Fig.~\ref{Fig.14}(a)) demands case-by-case analysis of the
QCD--hadron relations' singularity structure \cite{LMS8}; the
reward for the peeling efforts is a QCD sum rule closer to one's
targeted tetraquarks (Fig.~\ref{Fig.14}(b)).

In retrospect, for both possible quark-flavour distributions it is
achievable to devise formulations of QCD sum rules
\cite{LMS6,LMS8} that enable users to zoom in on (peculiarities
of) multiquarks.
\end{multicols}\begin{multicols}{2}

\section{Concise summary, conclusion, and prospect}We showed, for
flavour-exotic tetraquarks, how to construct a variant of QCD sum
rule for multiquarks not burdened by any contribution bearing no
relationship to multiquarks \cite{LMS6,LMS8}, and we don't expect
to encounter unsurmountable obstacles in the course of extending
this concept to other multiquark~varieties.

\section*{Acknowledgements}D.~M.\ acknowledges support from the
Austrian Science Fund (FWF), Grant No.~P29028; H.~S.\ is grateful
for support by the EU research and innovation programme Horizon
2020, Grant No.~824093; both D.~M.\ and H.~S. thank for obtaining
support from joint CNRS/RFBR grant No.~PRC Russia/19-52-15022.

\end{multicols}\clearpage\medline\begin{multicols}{2}

\end{multicols}
\end{document}